\documentclass{kluwer}    

\newdisplay{guess}{Conjecture}

\begin{document}                                                                                   
\input psfig
\def\kms{km s$^{-1}$}

\begin{article}
\begin{opening}         
\title{Dark-Matter Distribution inferred
from High-accuracy Rotation Curves } 
\author{Yoshiaki \surname{Sofue}}  
\runningauthor{Yoshiaki Sofue}
\runningtitle{Dark Matter Distributions}
\institute{Institute of Astronomy, University of Tokyo, Mitaka, Tokyo 181-8588, Japan}
\date{Sep 18 1999}

\begin{abstract}
Using high-accuracy rotation curves of spiral galaxies, we
derive distributions of the surface mass density.
Comparing with the luminosity profiles, we show that
the dark mass in disk and bulge distributions follows those of 
luminous (stellar) mass, while luminous mass in halos does not follow dark mass.
The dark-mass fraction (DMF) increases with the radius in the disk,
and rapidly toward edge in the halo.
In some galaxies, DMF increases  toward the center,
indicating a massive dark core.
\end{abstract}

\end{opening}           

\section{Introduction}

Dark mass dominates the mass of galaxies
(e.g., Rubin et al  1985; Kent 1987; Forbes 1992; Persic et al 1996).
On the other hand, the total mass of galaxies
is approximately proportional to the luminous mass
(Burstein et al 1997).
This indicates that the dark mass follows the luminous mass in a large scale.
A question may arise, whether the dark mass
in individual galaxies locally follows the luminosity mass, or
they are decoupled from each other.
This may be answered by mapping the dark mass in individual
galaxies.
We have shown that the CO molecular line is useful to derive
accurate rotation curves, because of its high
concentration in the center as well as for its negligible extinction
(Sofue 1996, Sofue et al 1997, 1998, 1999).
Recent high-dynamic-range CCD spectroscopy in optical lines
has also made it possible to obtain high accuracy rotation curves
(Rubin et al 1997; Sofue et al 1998).
In this paper, we discuss the general properties of
high-accuracy rotation curves, and use them to
derive distributions of surface mass density, 
mass-to-luminosity ratio, and the dark mass fraction.
Based on these, we discuss the correlation of dark and luminous mass in galaxies.

\section{General Properties of Rotation Curves}

In Fig. 1 we show well-sampled rotation curves obtained by
combining CO, CCD H$\alpha$, and HI observations.
We may summarize the universal properties of  rotation curves
as follows.

{\parindent=0pt
(1)Massive galaxies show very steep rise of
rotation in the central region.
The rotation velocity often starts from a finite value at the center.
However, small-mass galaxies tend to show gentler rise.

(2) Steep central peaks and/or shoulders corresponding to bulges;

(3) Broad maximum in the disk; and

(4) Flat rotation toward the edge.
}

\begin{figure}
\psfig{figure=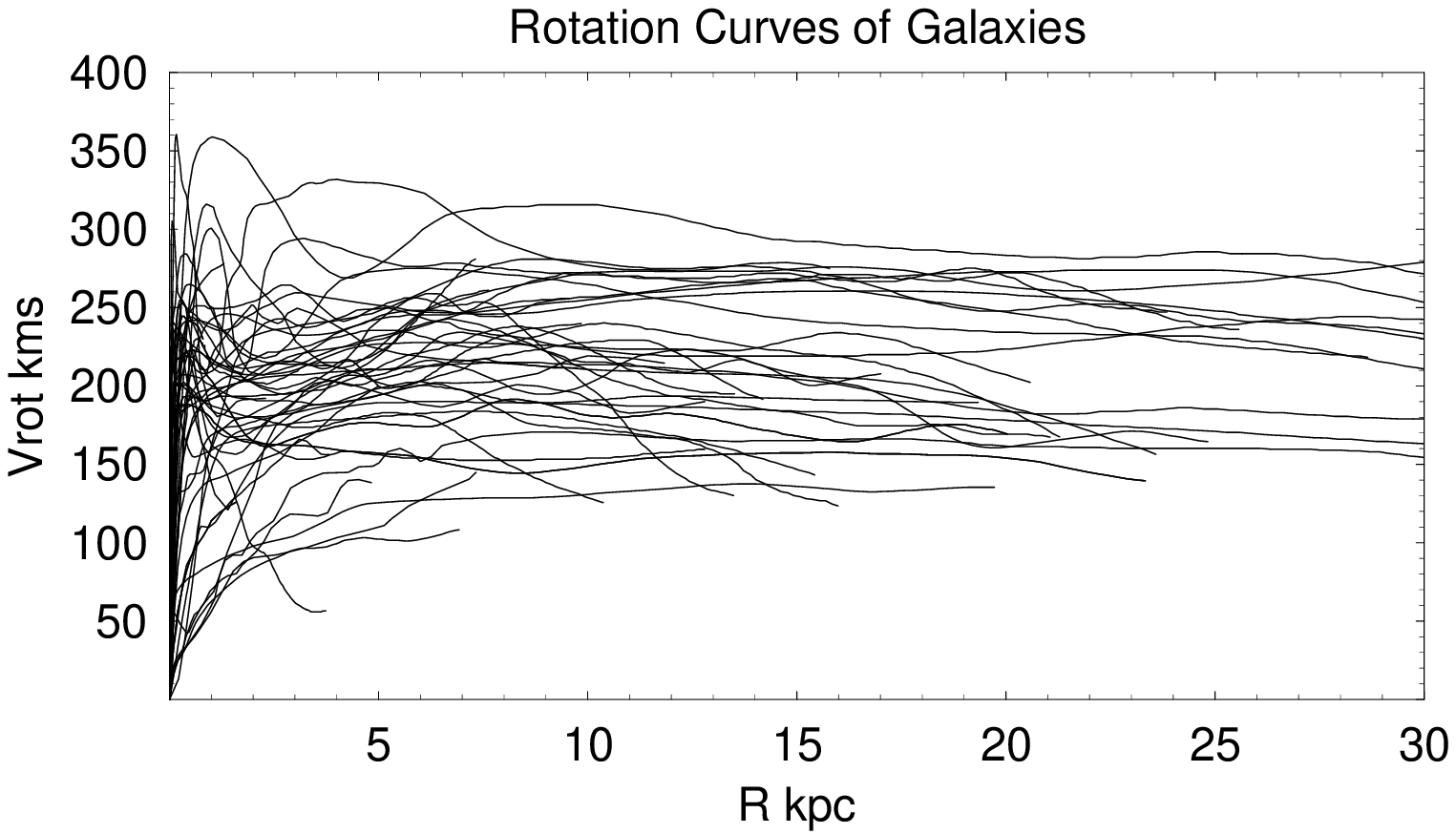,height=3cm}
Fig. 1. High-accuracy rotation curves  of Sb, Sc, SBb
and SBc galaxies.
\end{figure}

\begin{figure}
\psfig{figure=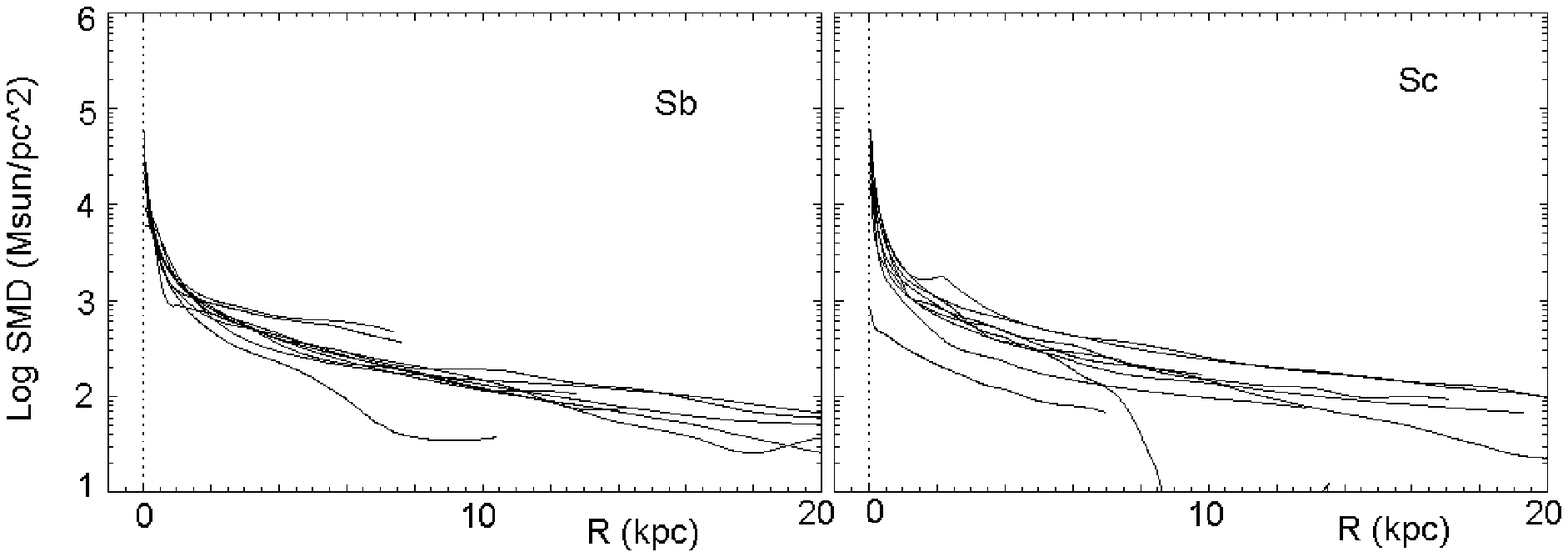,height=2.5cm}
Fig. 2.  Surface-mass density profile for Sb and
 Sc galaxies (Takamiya and Sofue 1999).
\end{figure}

\section{Surface-Mass Density}

We have developed a method to derive a differential surface mass
density as a function of  radius in order to compare the mass distribution
with observed profiles of the surface luminosity, once an accurate
rotation curve of a galaxy is given (Takamiya and Sofue 1999).
We assume that the 'true' mass distribution in a real disk galaxy will
be sandwiched by calculated spherical and flat-disk distributions.
The obtained SMD profiles are shown in Fig. 2.
Sb and Sc galaxies show similar surface-mass distributions.
In the disk region at radii 3 to 10 kpc the SMD decreases
exponentially outward.
In the bulge region, some galaxies show steep increase of SMD toward the center,
indicating high density cores.
The LMC shows a similar mass distribution as inferred
from an analysis of the HI line data obtained by Kim et al (1998)
(Sofue 1999).

\begin{figure}
\psfig{figure=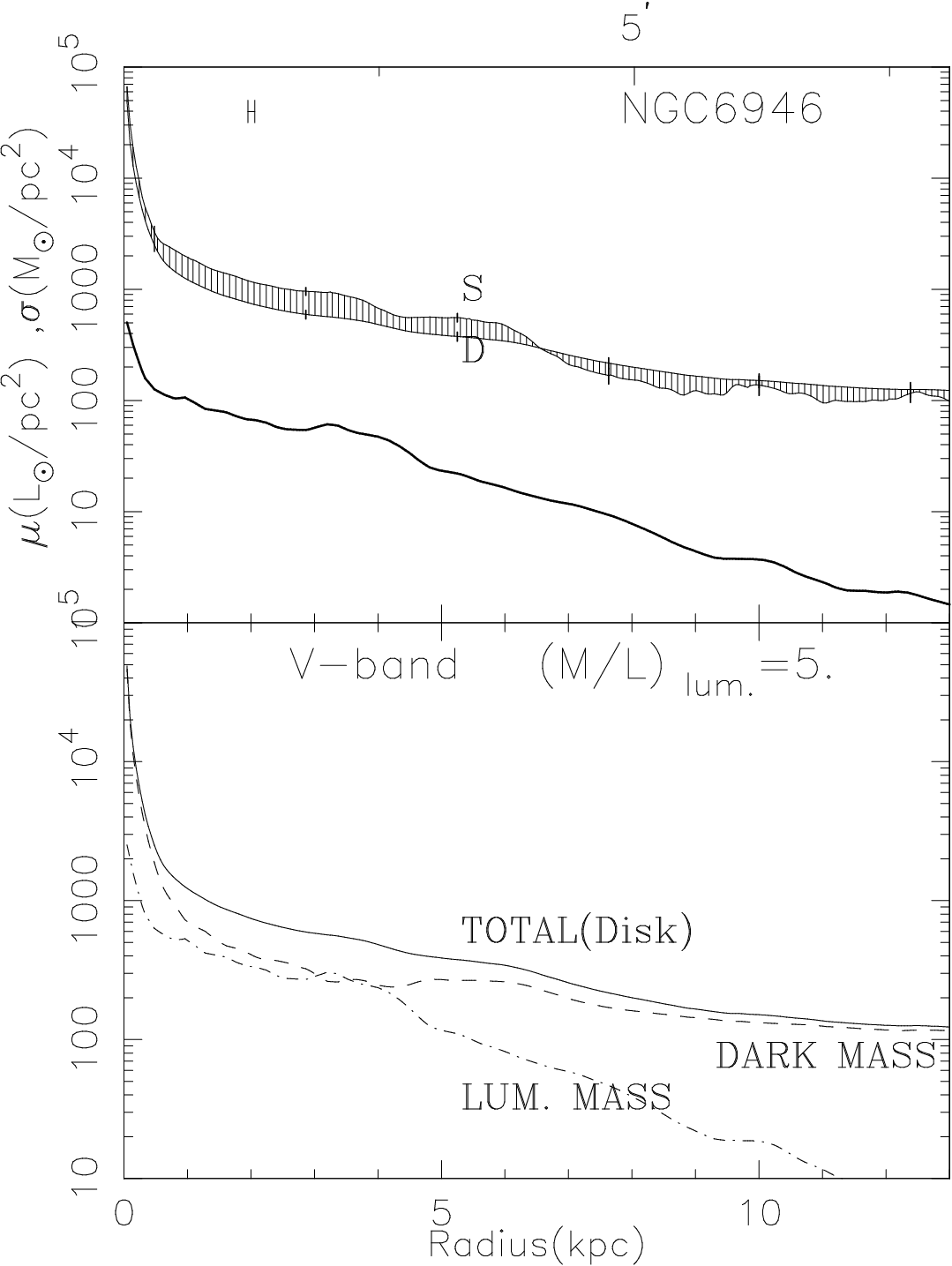,height=5cm}
Fig. 3. Dynamical mass and luminosity profiles (upper panel),
and dark and luminous masses (lower panel) for
NGC 6946 (Takamiya and Sofue 1999).
\end{figure}

\section{Correlation of Dark and Luminous Mass Distributions}

The surface mass density can be directly compared with
observed surface luminosity, from which we can derive the
mass-to-luminosity ratio (M/L).
By subtracting the luminous mass we can, then, derive the
distribution of dark mass.
Fig. 3 shows a typical case for NGC 6946, which was found to be
rather general for many other galaxies.
The general correlation between luminous mass and dark mass can be
summarized as follows.

{\parindent=0pt
(1) Luminous mass (stars) in disk and bulge
approximately follows the dark mass, and vice versa.

(2) The dark mass fraction increases with the radius in
the outer disk.

(3) The total dark mass and total luminosity of
galaxies are correlated as a whole.

(3) However, luminous matter in the halo (e.g., 
beyond $\sim 5$ disk scale radii) does not follow dark mass.
}

\begin{figure}
\psfig{figure=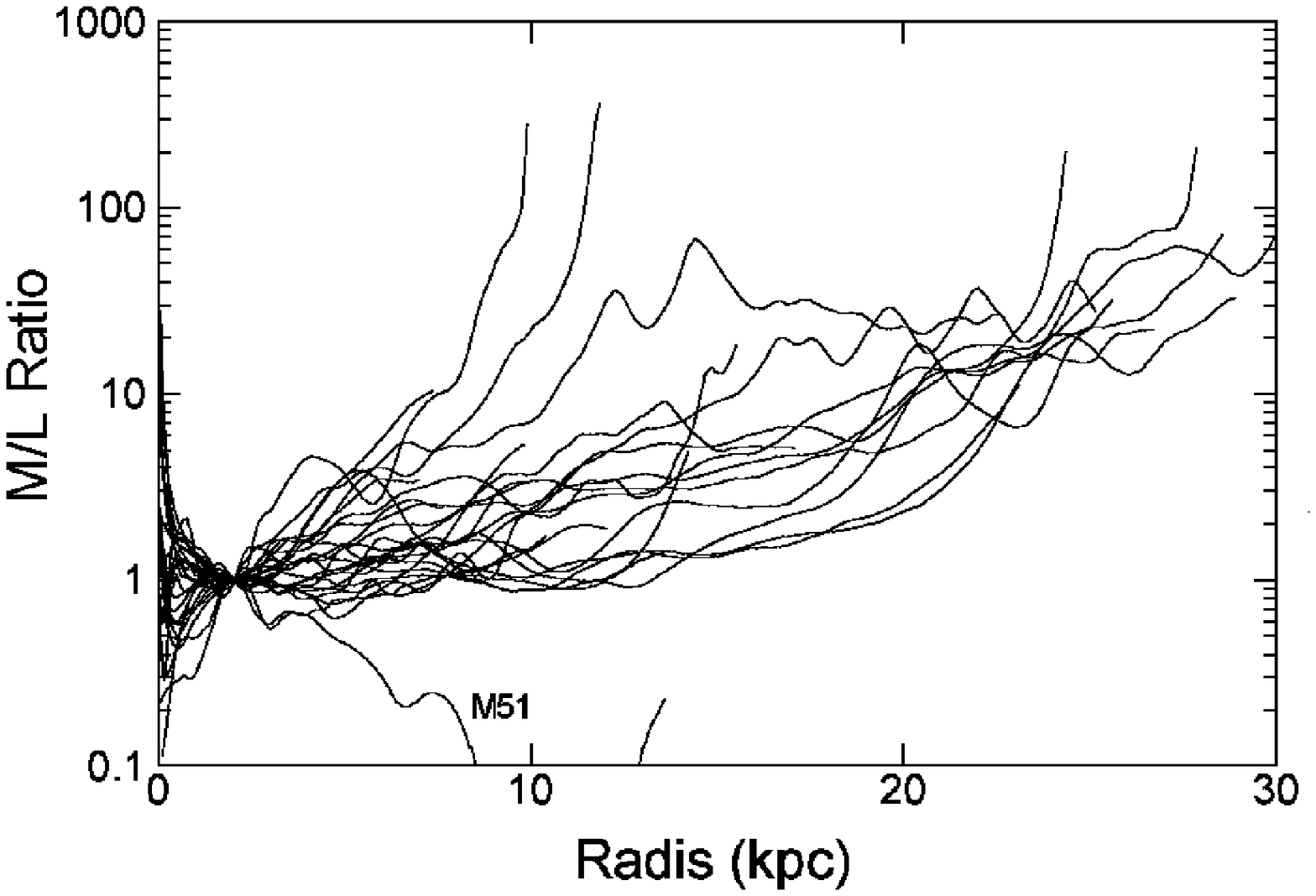,height=3cm}
\vskip -3cm
\hskip 5.5cm
\psfig{figure=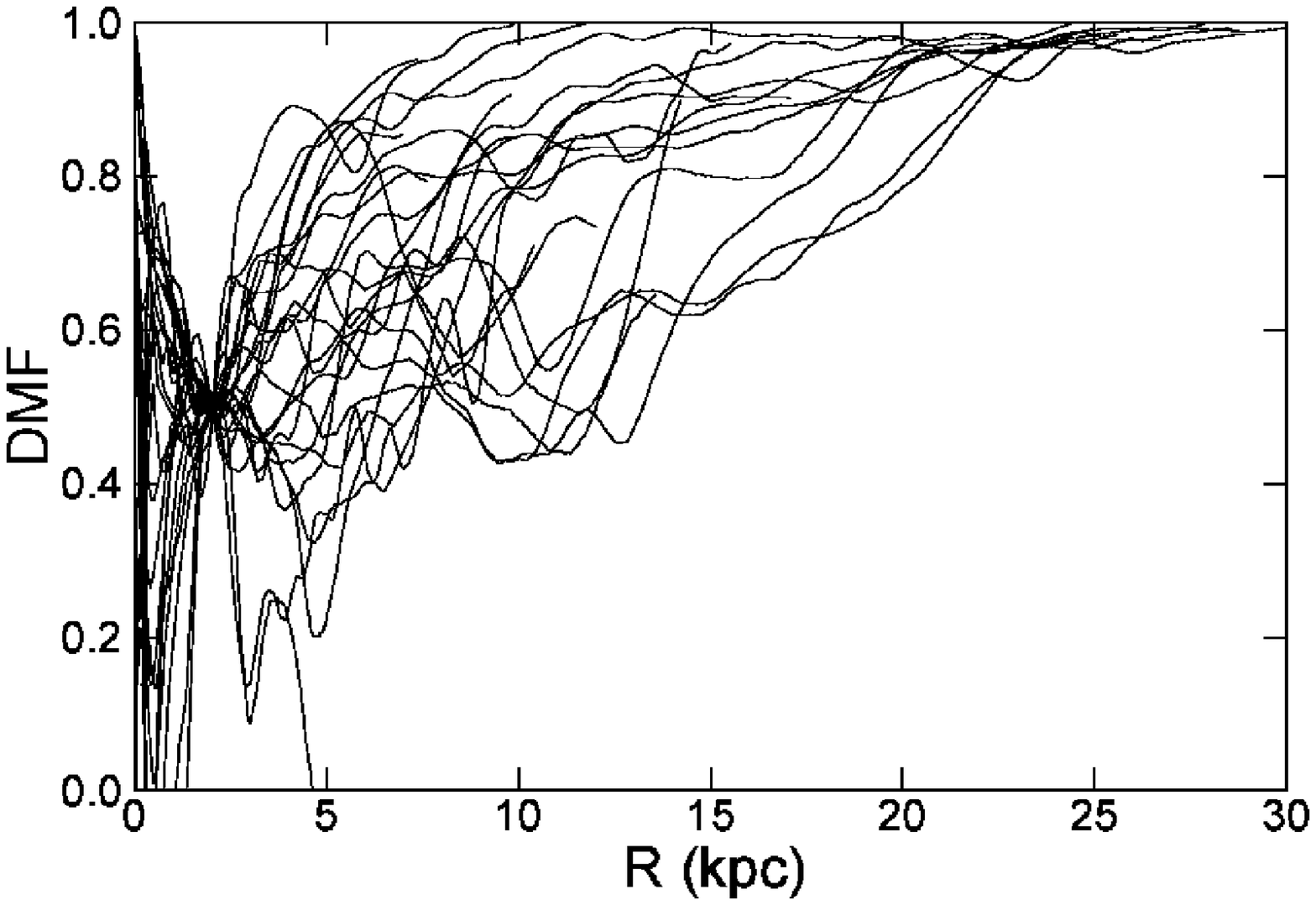,height=3cm}

Fig. 4.  M/L ratio (left panel: Takamiya and Sofue 1999)
and dark-mass fraction for Sb and Sc galaxies
\end{figure}

\section{Mass-to-Luminosity Ratio and Dark-Mass Fraction}

Radial variations of the M/L and 
dark-mass fraction are shown in Fig. 4 as normalized at $R=2$ kpc
(Takamiya and Sofue 1999).
We have defined a quantity DMF by $1 - a/(M/L)$
with $a$ a constant and M/L being normalized at $R=2$ kpc.
Although the dark mass and luminosity are correlated in the disk and bulge
in general, the detailed distribution of M/L and DMF indicates that they
are not constant, but vary  significantly within a galaxy.
M/L and DMF behavior in spiral galaxies may be summarized as follows.
\vskip 1mm

{\parindent=0pt
(1) M/L and DMF vary drastically within the central bulge.
In some galaxies, it increases toward the center,
suggesting a dark massive core.

(2) M/L and DMF gradually increase in the disk region,
and the gradient increases with the radius.

(3) M/L and DMF increase drastically toward
the outer edge, indicating massive dark halo.
}


\end{article}
\end{document}